\documentstyle[10pt,aaspp4,flushrt,tighten]{article}

\lefthead{Statler et al.}
\righthead{Intrinsic Shapes of Elliptical Galaxies. IV.}

\begin{document}

\title{Uncovering the Intrinsic Shapes of Elliptical Galaxies.
\\IV. Tests on Simulated Merger Remnants}

\author{Thomas S. Statler\altaffilmark{1},
Heath Lambright,
and 
Jakob Bak}
\affil{Department of Physics and Astronomy, 251B Clippinger Research
Laboratories, Ohio University, Athens, OH 45701, USA}
\altaffiltext{1}{tss@helios.phy.ohiou.edu}

\vskip -1.9in {\hfill \sl Astrophysical Journal, accepted}
\vskip 1.8in

\begin{abstract}

We test the methods developed in previous papers for inferring the
intrinsic shapes of elliptical galaxies, using simulated objects from
$N$-body experiments. The shapes of individual objects are correctly
reproduced to within the statistical errors; close inspection of the results
indicates a small systematic bias in the sense of underestimating both
the triaxiality and short-to-long axis ratio. We also test
the estimation of parent shape distributions, using samples of
independent, randomly oriented objects. The best results are statistically
accurate, but on average the parent distributions are again
slightly biased in the same sense. Using the posterior
probability densities for the individual objects, we estimate the
magnitude of the bias to be $\lesssim 0.1$ in both shape parameters. The
results support the continued use of these methods on real systems.

\end{abstract}


\section{Introduction\label{s.introduction}}

Observational estimates of the true three-dimensional shapes of elliptical
galaxies serve as diagnostics of the physics of galaxy formation and
evolution. Numerical simulations of protogalactic collapse (Dubinski \&
Carlberg \markcite{DC91}1991), mergers (Weil \& Hernquist \markcite{WH96}1996,
Barnes \& Hernquist \markcite{BH98}1998), and black-hole growth
(Merritt \& Fridman \markcite{MF98}1996, Merritt \& Quinlan
\markcite{MQ98}1998) all suggest that dark halos {\it should\/} tend
toward axisymmetry with increasing dissipation or central mass concentrations.
But whether this has, in fact, happened is difficult to establish.
The effort to uncover the intrinsic shape distribution has a long history
(reviewed by Statler \markcite{Sta95}1995, updated by Bak \& Statler
\markcite{BS00}2000, hereafter BS). Photometric methods, introduced by
Hubble \markcite{Hub26}(1926) and revisited by many authors (most recently
Ryden \markcite{Ryd92}1992, \markcite{Ryd96}1996; Fasano \markcite{Fas95}1995),
have been effective in constraining the distribution of flattenings,
but not in establishing the relative numbers of triaxial and axisymmetric
systems. For constraining triaxiality, dynamical information is
necessary, with both gas (Bertola et al. \markcite{Ber91}1991) and stellar
(Binney \markcite{Bin85}1985, Franx et al. \markcite{FIZ91}1991, Tenjes et al.
\markcite{Ten93}1993) kinematic data having been used for this purpose.

Earlier papers in this series (Statler \markcite{S94a}1994a, Statler
\& Fry \markcite{SF94}1994, Statler \markcite{S94b}1994b, hereafter
Papers I, II, and III respectively) presented a
method for statistically estimating the intrinsic shapes of
elliptical galaxies from observations of their radial velocity fields
and surface brightness distributions. Paper III included
a test of the method on the end product of an $N$-body collapse simulation
(Dubinski \markcite{D92}1992). Using simulated observations of the
object along a single line of sight, the method could
recover the true shape only to within a $2\sigma$ error region; this
result was attributed to an unfortunate line of sight. Since then,
the method has profitably been applied to real systems, to derive both the
shapes of individual galaxies (NGC 3379, Statler \markcite{S94c}1994c,
Statler \markcite{S00}2001, NGC 1700, Statler et
al.\ \markcite{SDS99}1999, hereafter SDS)
and the parent shape distribution of a sample
(Davies \& Birkinshaw \markcite{DB88}1998) of radio ellipticals
(\markcite{BS00}BS). Nonetheless, there
remains a lingering impression that the performance on realistic simulated
systems is ``disappointing'' (Binney \& Merrifield \markcite{BM98}1998).

In this paper we present more extensive tests, using a
homogeneous set of simulated elliptical galaxies from group-merger
experiments (Weil \& Hernquist \markcite{WH94}1994,
\markcite{WH96}1996, hereafter WH94 and WH96, collectively WH).
We demonstrate that the method performs well, both
in estimating the shapes of individual objects and in deriving the shape
distribution of small samples. The presentation is ordered as follows:
Section 2 briefly describes the test objects and the methods employed, with
references to earlier papers where complete expositions can be found. Section
3 describes the tests performed and their results. Section 4 deals with
residual systematic bias and its possible origin, and Section 5 sums up.

\section{Methods\label{s.methods}}

\subsection{Simulated Data\label{s.data}}

\subsubsection{Merger Remnants\label{s.simulations}}

The simulated ellipticals used for this study are late-time snapshots of
the group-merger simulations of WH, listed as 1--6 in Table 1 of
WH96. The simulations were performed using a tree code, with
393216 particles representing luminous matter
and an equal number representing dark matter, initially distributed
as six disk galaxies with dark halos. The disk/halo mass ratio was
$1:5.8$; in case 5, the
galaxies also contained bulges one-third the mass of the disk.
The initial systems had identical macroscopic properties, except in case
6, where the masses of two systems were doubled.
The six progenitors were distributed inside a sphere with a radius 30 times the
disk scale length, and were given center of mass velocities drawn from
a Maxwellian distribution.
In five of the six cases, all group members
merged into a single ellipsoidal remnant; case 4 ejected one
galaxy at an early time.

\subsubsection{Observations\label{s.observing}}

Each of the merger remnants may be rotated to any orientation.
To keep the nomenclature unambiguous, we will refer to the six remnants as
the {\it true objects\/}, and to a particular remnant in a particular
orientation as an {\it observed object\/}.

For each observed object we project the phase space particle distribution
for the luminous matter along the line of sight and create
photometric and kinematic data sets from the zeroth and first velocity
moments, using a software pipeline developed by Bak \markcite{B00}(2000).
The projected distribution is first binned onto a tree-like
hierarchical grid, and the surface brightnesses and mean radial velocities
smoothed and interpolated onto a uniform grid using thin-plate splines.
The resulting surface brightness image is analyzed photometrically using
the ELLIPSE task in IRAF/STSDAS, which outputs radial profiles of
surface brightness, ellipticity, and major axis position angle (PA).
Based on a radial average of
the latter, 1-dimensional cuts of the radial velocity field (i.e.,
rotation curves) are extracted along the major and minor axes and at
diagonal position angles $\pm 45\arcdeg$ from the major axis; while the actual
PA sampling in published data varies widely, this choice
reflects the sampling in recent observations used with this approach
(Statler et al.\ \markcite{SSC}1996, Statler \& Smecker-Hane
\markcite{SS99}1999). In some
cases with significant isophotal twists, the velocity field is resampled
at position angles keyed to the average major axis PA in the range of
radii where the data are used (see \ref{s.averages} below).\footnote{In
two cases kinematic cuts were erroneously made at PAs unrelated to the
photometric axes. Since the $45\arcdeg$ spacing of the cuts was preserved
and the actual sampled PAs were correctly propagated through the modeling
procedure, we left the error uncorrected.}

\subsubsection{Radial Averages\label{s.averages}}

Use of the observational material exactly parallels that of Statler
\markcite{S94c}(1994c) and BS, where the goal is to produce estimates of
the mean shape in the main body of the galaxy. Because projection of the
model velocity fields is robust only where the rotation curve is not
steeply rising, we exclude data at small radii. Each velocity profile is folded
about the center and averaged, assuming antisymmetry. For each observed object
we estimate by eye the radius at which the largest-amplitude rotation
curve flattens and omit the data inside this point. We also discard data
beyond approximately 4 effective radii since data past this point are
unavailable for real galaxies. Typically the data we retain span a
factor of 3 to 5 in radius (e.g., $0.8$ to 3 effective radii).
We take an unweighted average of the radial
velocities on each PA between the inner and outer radii, and the
uncertainty associated with the average is taken to be the
unweighted standard deviation. Photometric data are handled similarly;
we adopt error weighted averages of the major-axis position angle and
ellipticity over the same radial interval, using standard deviations
as the uncertainties. Each observed object is hence described by 4 radially
averaged velocities and one mean ellipticity.

\subsection{Dynamical Modeling\label{s.modeling}}

\subsubsection{Models\label{s.models}}

The dynamical models are presented in Paper I and their use discussed
in subsequent papers. The models assume a stationary potential, with
observed rotation arising from a mean streaming of the ``stellar fluid''
in the nonrotating figure. The total internal velocity field is
the vector sum of mutually crossing flows generated by short-axis and
long-axis tube orbits, each of which is assumed to follow confocal
streamlines. Other orbit families have zero mean motion and contribute only
to the density.
The shapes of the streamlines are linked to the triaxiality
of the total mass distribution (Anderson \& Statler \markcite{AS98}1998).
Given ellipsoidal distributions for the
luminous and total mass, the velocity field is calculated by solving the
equation of continuity subject to appropriate boundary conditions. The
boundary conditions are described by a number of adjustable parameters,
allowing a broad variety of dynamical configurations.
The models are then projected along the line of sight assuming that the
luminosity density and mean rotation vary as power laws with radius.

\subsubsection{Bayesian Shape Estimate\label{s.bayes}}

The Bayesian methodology for estimating the shapes of single objects
is discussed in detail in Paper III. For
each observed object, a lengthy exploration of the parameter space (typically
$\sim 10^7$ individual models) results in a multidimensional likelihood
function $L(T,c_L,\Omega,\mbox{\boldmath $d$})$, where $T$ is the
triaxiality of the mass distribution, $c_L$ is the short-to-long axis
ratio of the luminosity distribution, and $\Omega$ and $\mbox{\boldmath $d$}$
represent the orientation and the remaining dynamical parameters, respectively.
Models are computed over the same grid of parameters used by \markcite{BS00}BS,
to match previous applications to real data (see \S\ 2.4 of BS, \S\ 2.1 of
SDS, \S\S\ 2 and 4.1 of Statler 1994c for details).
We normally assume a flat prior distribution in all
parameters; the posterior probability density $P(T,c_L)$, which constitutes
the estimate of intrinsic shape, is then obtained by integrating
$L(T,c_L,\Omega,\mbox{\boldmath $d$})$ over the nuisance parameters
$\Omega$ and $\mbox{\boldmath $d$}$ and normalizing the result. The flat
distribution in $\mbox{\boldmath $d$}$ represents ``maximal ignorance''
of the true dynamical configuration. Other prior assumptions may be
accommodated by limiting the integration over $\mbox{\boldmath $d$}$ to
a subregion of parameter space; a relevant example would be an assumption
that rotation were purely around the intrinsic short axis.

\subsubsection{Parent Distribution\label{s.parent}}

Given a sample of observed objects, we can also attempt to recover the
parent shape distribution for the set of true objects. This
proceeds according to the methods described by BS. We assume that the
sample is randomly oriented, and start with a flat model for the parent
distribution, $F(T,c_L) = {\rm const}$. The posterior probability density
$P_i(T,c_L)$ for each observed object $i$ is given by the normalized product
of $F$ with the integrated likelihood $L_i(T,c_L)$. (Initially, $P_i = L_i$.)
The $P_i$'s are summed, and the sum is smoothed with a nonparametric
smoothing spline and normalized. This yields an improved estimate for
$F$, which is multiplied by the $L_i$'s and iteratively cycled through
the same procedure. The smoothing parameter in the spline is chosen at
the first iteration to optimize a cross-validation score,\footnote{The
likelihood cross-validation score (Coakley \markcite{Coa91}1991) is the
average over $i$ of the log-likelihood of
the $i$th observation in the model fitted to all {\it but\/} the $i$th
observation (BS, \S\ 2.3, equations [10] and [11]); thus an optimized score
gives the model that is most likely to correctly predict the next observation.
The use of cross-validation with smoothing splines is
discussed fully by Green \& Silverman \markcite{GS94}(1994).}\ and we
conservatively stop iterating when the maximum fractional change per
iteration in $F$ drops below 10\%. Once the final parent distribution is
obtained, its normalized product with the integrated likelihoods gives
corrected shape estimates for the individual objects.

\subsection{True Shapes\label{s.shapes}}

The axis ratios of the true objects are computed directly from the
particle distributions. The values tabulated by WH96 are not suited to
our purposes, as they represent averages over subsets of particles sorted
by binding energy. We need instead the axis ratios of the isodensity
surfaces, averaged over the range of radii from which the kinematic data
are drawn. We locate the principal axes and measure the axis ratios by
diagonalizing the inertia tensors of thin ellipsoidal shells about a
fixed center. The shell thicknesses are adjusted to maintain
approximately 2000 particles in each shell, and the axis ratios are
manipulated until the eigenvalues of the inertia tensor match those for
a constant density shell of the same shape. 

We find that all of the objects are highly triaxial in the inner parts,
becoming nearly axisymmetric at moderate radii. The direction of the
minor axis stays, for the most part, well aligned through each system;
however, as the middle-to-long axis ratio $b$ rises above
approximately $0.85$, the long-axis direction in the plane normal to
the short axis becomes ill defined and wanders from one shell to the next.
While we can compute the radial average of $c$ by elementary methods,
a similar average of $b$ would result in an overestimate of the
triaxiality, since particle noise will always make a round contour elliptical.
Consequently we adopt a complex representation of the ellipticity in the
equatorial plane, ${\cal E} = \epsilon\, e^{2i\phi}$, where $\epsilon =
1-b$ and $\phi$ is the long-axis position angle. We compute the radial
average $\langle {\cal E} \rangle$, from which the average axis ratio is
$\langle b \rangle =1-|\langle {\cal E}\rangle|$, and the triaxiality
follows according to $\langle T \rangle = (1 - \langle b \rangle^2) /
(1 - \langle c \rangle^2)$.

\begin{figure}[t]{\epsfxsize=4.5in\hfil\epsfbox{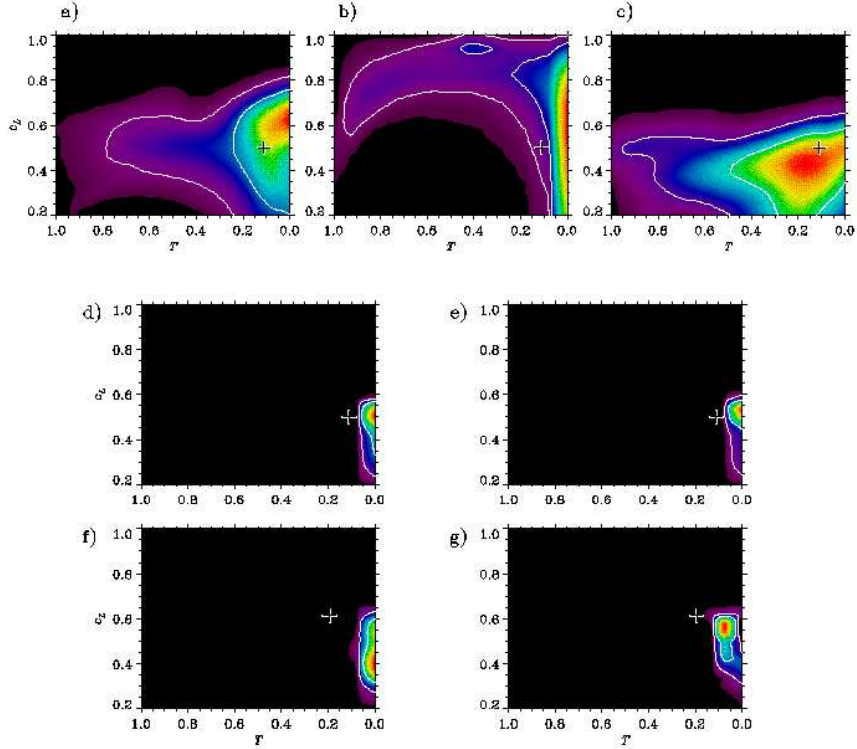}}
\caption{\footnotesize
Probability densities for the shapes of single objects, in the space of
total-mass triaxiality $T$ and luminous short-to-long axis ratio $c_L$.
Contours show the 68\% and 95\% highest posterior density (HPD) regions,
analogous to $1\sigma$ and $2\sigma$ error ellipses. Crosses show the
true shapes, averaged over the radial interval from which the data are
drawn. ($a$--$c$) Typical results for 3 different views of WH's remnant 2,
using all models (the ``maximal ignorance'' result). ($d$) 11 views
of remnant 2, modeled with maximal ignorance of the dynamics but combined
with the prior knowledge that they are independent observations of the same
object. ($e$) Same as ($d$), using only models with rotation aligned
with the intrinsic short axis. ($f$,$g$) As in ($d$,$e$), but for
remnant 1.
\label{f.single}}
\end{figure}

\section{Statistical Tests\label{s.tests}}

\subsection{Single Objects\label{s.single}}

To test the method's ability to recover the shape of an individual system,
we project remnant number 2 along 11 different lines of sight uniformly
distributed over one hemisphere, creating 11 observed objects from
one true object. We initially model the observed objects independently,
as if they were different systems. The top row ($a$--$c$)
of Figure \ref{f.single} shows posterior probability densities obtained
in the ``maximal ignorance'' case for three of the objects.
Contours in each panel show the 68\% and 95\% highest posterior density
(HPD) regions, analogous to $1\sigma$ and $2\sigma$ error ellipses.
These results are similar to those for real
systems, in that nearly oblate shapes are preferred, but highly triaxial
ones are not excluded. Crosses show the shape of the true object, where
we have measured $T$ and $c_L$ separately from the total mass and
luminous mass distributions, respectively, averaged over the intervals
from which the data are drawn. As in
previous work, we find that the inferred shape and the accuracy of the
result depend on the particular line of sight; it {\it is\/} possible to be
fooled by an unlucky view. However, in this test the correct shape falls
outside the 68\% highest posterior density (HPD) region only twice out
of 11 trials.

\begin{figure}[t]{\epsfxsize=4.5in\hfil\epsfbox{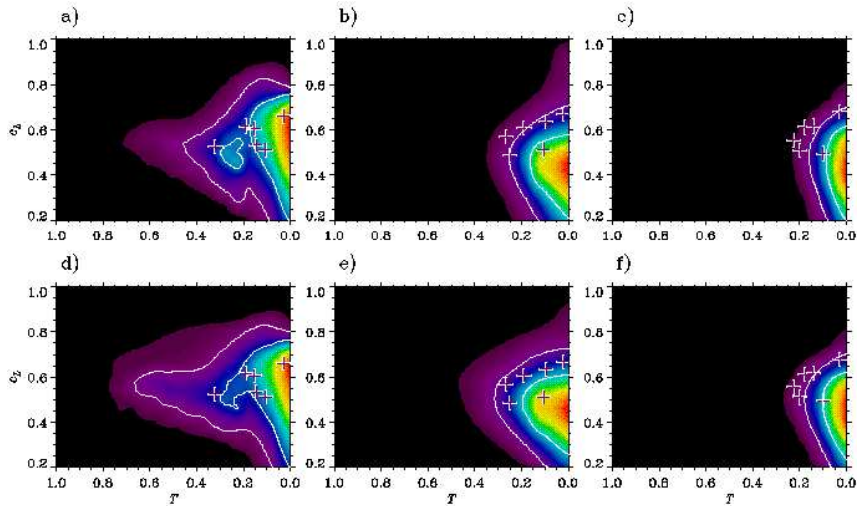}}
\caption{\footnotesize
Parent shape distributions derived from 3 samples of the 6 WH remnants
oriented at random: ($a$,$d$) Sample 1; ($b$,$e$) Sample 2; ($c$,$f$)
Sample 3. Results are shown in the {\em top row\/} for the maximal ignorance
case, and in the {\em bottom row\/} for aligned rotation. Crosses show true
shapes.
\label{f.parents}}
\end{figure}

To check for systematic bias, we exploit our knowledge that the 11
observed objects in fact have the same true shape. The proper
probability distribution in this case is the normalized product of the 11
separate distributions $P_i(T,c_L)$ (Paper III, \S\ 3.2).
The result is shown in Figure
\ref{f.single}d. Despite the width of the individual $P_i$'s, the
product is a sharp spike, with its peak at $(T,c_L)=(0,0.5)$. This
reproduces the true shape to within the resolution of the model grid in
$c_L$, and to within approximately $0.1$ in $T$. We know, however, that
the maximal ignorance case does not correctly describe the dynamics of the true
objects. WH94 report that all of their group merger remnants have
angular momenta closely aligned with the short axes. Therefore, we also
perform the calculation using only the models that rotate about their
short axes; the result is shown in Figure \ref{f.single}e. For this
particular object, the change makes little difference.
The bottom row of Figure \ref{f.single} shows results for the same tests
repeated on remnant number 1, for the maximal ignorance and aligned
rotation cases. The results are not quite as good for this object,
though now there is improvement with the correct assumption of aligned
rotation. These models correctly detect the nonzero triaxiality,
but underestimate both $T$ and $c_L$ by of order $0.1$.

While the above tests suggest a systematic bias in the shape estimates, the
bias is small compared to typical statistical errors. The performance of the
individual $P_i$'s for the 22 observed objects taken together is close to
statistically perfect. For both maximal ignorance and aligned rotation,
the true shapes fall inside the 68\% HPD region 13 times (15 expected)
and inside the 95\% HPD region 21 times (21 expected).

\begin{figure}[t]{\epsfxsize=4.5in\hfil\epsfbox{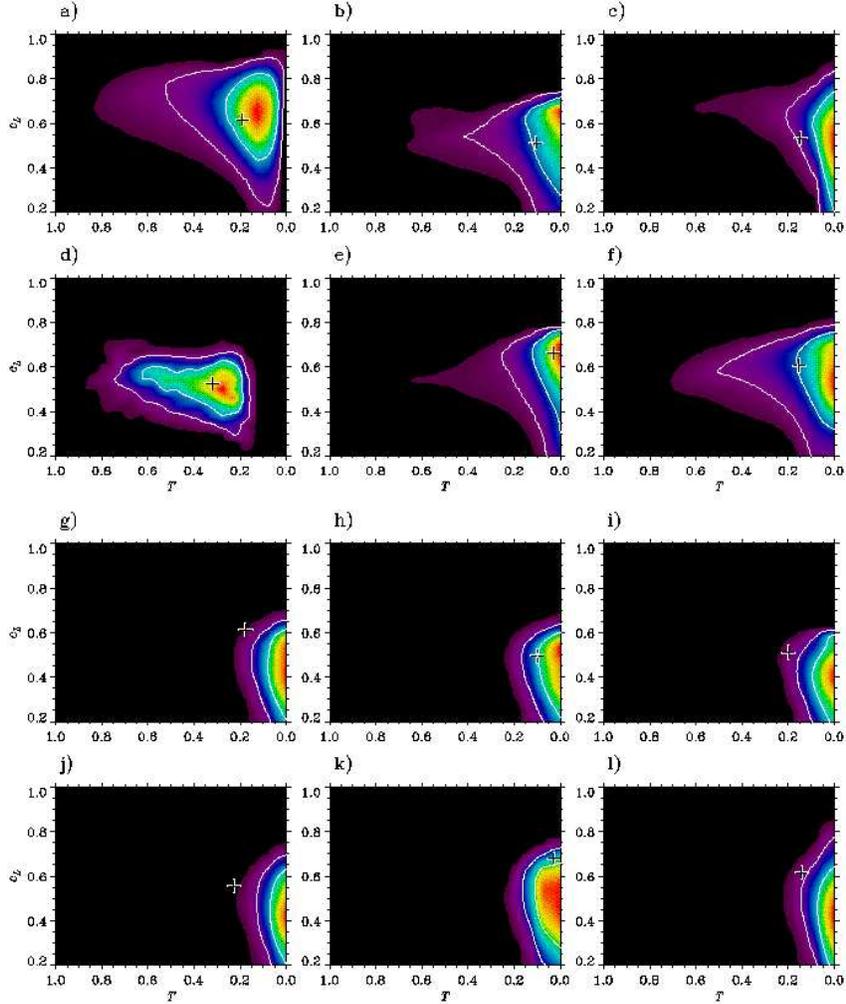}}
\caption{\footnotesize
Posterior probability distributions for the shapes of individual objects
in the samples, using the parent distributions of
Fig.\ \protect\ref{f.parents}. Crosses show true shapes. ($a$--$f$)
Results for remnants 1--6 in the best case (Sample 1, aligned rotation).
($g$--$l$) Same as ($a$--$f$) but for the worst case (Sample 3, maximal
ignorance).
\label{f.individuals}}
\end{figure}

\subsection{Simulated Samples\label{s.samples}}

To test the estimation of parent shape distributions, we construct three
simulated samples, each consisting of the 6 true objects oriented at random.
The maximal-ignorance parent distributions obtained for the three samples
are shown in the top row of Figure \ref{f.parents}; the corresponding
results using only models with rotation aligned with the short axis
appear in the bottom row. Crosses show the true mean shapes, which vary
slightly between samples because the means are computed over slightly
different radial intervals. The samples are ordered by the quality of
the results. The maximal ignorance distribution for Sample 1 successfully
puts 4 objects inside the 68\% HPD contour, and all 6 inside the 95\%
contour. The aligned rotation models perform slightly better in all
cases, moving one more object inside the 68\% HPD region in Sample 1.
In the worst case, the maximal ignorance distribution for Sample 3, only
2 objects fall within the 95\% HPD region and none within the 68\% region.

The posterior shape estimates for the individual objects can be used to
address systematic bias. These are shown for the best and worst cases in
Figure \ref{f.individuals}. The aligned rotation models for Sample 1
correctly detect the nonzero triaxiality of the two most triaxial
objects ($a$, $d$),
but give most-probable triaxialities of zero for the other four.
The objects of Sample 3 are predicted to be more axisymmetric
and flatter than they actually are; similar results are seen for Sample 2.
Taken as a group, the 18 observed objects in the three samples fall
within their respective 68\% HPD regions 33\% of the time, and inside
the 95\% HPD regions 61\% of the time in the maximal ignorance case; the
corresponding percentages for the aligned rotation models are 44\% and
72\%. Shifting the true shapes uniformly by $-0.09$ in both $T$ and
$c_L$ raises these percentages to their expected values. We conclude that
systematic bias in the parent distributions is $<0.1$ in $T$ and $c_L$,
which is consistent in both magnitude and direction with the results of
section \ref{s.single} for single objects.

\section{Discussion\label{s.discussion}}

\begin{figure}[t]{\epsfxsize=4.5in\hfil\epsfbox{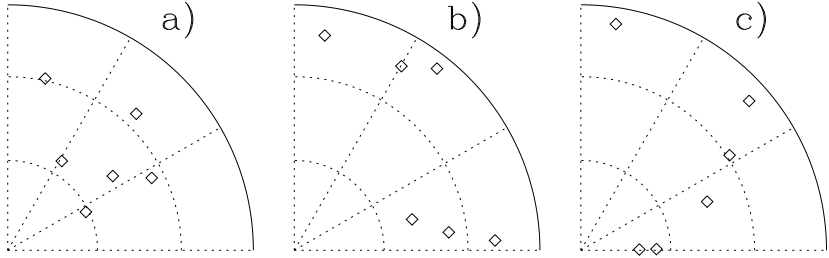}}
\caption{\footnotesize
Distribution of lines of sight for the observed objects in ($a$) Sample
1, ($b$) Sample 2, ($c$) Sample 3. In each panel the origin corresponds
to views down the short axis, the top edge to views down the long axis,
and the right edge to views down the middle axis. For these models views
are uniquely defined over only 1 octant, which is shown in equal-area
projection with dotted lines at $30\arcdeg$ intervals.
\label{f.orientations}}
\end{figure}

We would like to understand the origin of the small systematic bias in our
results. One possibility is that the small samples used in
\S\ \ref{s.samples} could be biased in orientation, but this is not the
case. Figure \ref{f.orientations} shows the distribution of the lines of sight,
relative to the principal axes of the observed objects, for each sample.
There is no obvious clustering of views. The isotropy of the distribution
can be quantified using $|Q_1|$, the largest (in absolute magnitude)
eigenvalue of the quadrupole moment tensor, normalized to the number of
objects. A Monte Carlo simulation of $10^5$ six-object samples gives a
skewed distribution with a mean of $0.44$, a mode of $0.36$,
and a standard deviation of $0.22$. The $|Q_1|$ values for Samples 1, 2,
and 3 are $0.35$, $0.42$, and $0.24$ respectively, so the samples are
neither abnormally anisotropic nor abnormally isotropic.

At any rate, bias in the 6-object samples, even if present, would not
account for the similar systematic error seen in the single-object tests of
\S\ \ref{s.single}. Unfortunately, we see no obvious reason why the method
should tend to underestimate the triaxiality and overestimate the flattening.
But we can speculate on three possibilities:

\begin{enumerate}

\item {\em Overly cautious use of the kinematic data.\/} The observed rotation
curves are averaged over a range of radii where they are well-behaved.
By indulging a natural tendency to exclude radii where the data
become ``weird,'' we may have reduced the signature of the very kinematic
asymmetries that are the hallmarks of triaxiality.

\item {\em Coarse dynamical grid.\/} \markcite{SDS99}SDS and \markcite{BS00}BS
emphasized the importance of the ``disklike'' or ``spheroid-like''
character of the internal velocity field to the inferred shape. The
model grid used here includes only the extreme cases on the assumption
that they should bracket the correct result. However, we occasionally encounter
situations where shapes inferred from intermediate models are more
triaxial than those from either extreme (Statler \markcite{S00}2001).

\item {\em Figure rotation.\/} Pattern speeds for the merger
remnants have not been measured. An addition of a solid-body
component to the streaming velocity field of a triaxial object can, in
principle, make it appear more axisymmetric than it actually is. This is a
long-standing unresolved issue, and there are still no dynamical models
for ellipsoidal systems that include figure rotation as an adjustable
parameter.

\end{enumerate}

\section{Summary\label{s.summary}}

We have tested the methods developed in previous papers for inferring the
intrinsic shapes of elliptical galaxies, using a set of six simulated
objects formed by group mergers of disk/halo systems. We have modeled
two of the objects over a uniform grid of orientations, and find that
their true shapes are recovered to within the statistical errors. A more
stringent test, using the prior knowledge that the same object is being
observed repeatedly, indicates a small systematic bias in the results, in
that both triaxiality and short-to-long axis ratio are underestimated by
roughly $0.1$. We have also constructed three simulated samples of 6
objects, and estimate the parent shape distribution in each case. The
result for one of the samples is statistically unimpeachable, with the
correct fractions of the sample falling within the 68\% and 95\% HPD
regions. The other two samples give results that are worse by an amount
consistent with the above bias.
Including prior knowledge of the internal dynamics of the objects
improves the results for all the samples. Examination of the posterior
shape estimates for the individual objects in the samples again suggests
a systematic bias $\lesssim 0.1$ in triaxiality and flattening, in the
same sense found in the earlier test. The source of this bias is not
understood, but may be related to the handling of the observational data, the
coarse grid of dynamical parameters in the models, or figure rotation
of the test objects. As a whole, these results support the use of our
methods to understand the nature of real elliptical galaxies.

\acknowledgments

We are grateful to Melinda Weil for providing data from her group merger
simulations and for a careful referee's report. This work was supported
by NSF CAREER grant AST-9703036.

\end{document}